\documentclass[12pt]{elsarticle}

\journal{High Energy Density Physics}

\usepackage{bm}
\usepackage{amsmath}
\usepackage{amssymb}
\usepackage{amsfonts}
\usepackage{units}
\usepackage{graphicx}
\usepackage[usenames]{color}
\usepackage{subfigure}
\usepackage{lipsum}
\biboptions{sort&compress}
\graphicspath{{../../dft/graphite/structure/}}

\newcommand{\beq}{\begin{equation}}
\newcommand{\eeq}{\end{equation}}
\newcommand{\bea}{\begin{eqnarray}}
\newcommand{\eea}{\end{eqnarray}}
\newcommand{\bal}{\begin{align}}
\newcommand{\eal}{\end{align}}

\newcommand{\gcc}{\,\unit{g/cm^{3}}}
\newcommand{\ev}{\,\unit{eV}}

\newcommand{\Mbar}{\,\unit{Mbar}}
\newcommand{\gpa}{\,\unit{GPa}}

\begin{document}

\title{The structure in warm dense carbon}

\author[HZDR]{J. Vorberger}

\author[hro]{K.U. Plageman}

\author[hro]{R. Redmer}

\address[HZDR]{Institut f\"ur Strahlenphysik, Helmholtz-Zentrum Dresden-Rossendorf e.V.,
	     01328 Dresden, Germany}

\address[hro]{Institut f\"ur Physik, Universit\"at Rostock, 18051 Rostock, Germany}

\date{\today}

%%%%%%%%%%%%%%%%%%%%%%%%%%%%%%%%%%%%%%%%%%%%%%%%%%%%%%%%%%%%%%%%%%%%%%%%%%%%%%%%
% Abstract
%%%%%%%%%%%%%%%%%%%%%%%%%%%%%%%%%%%%%%%%%%%%%%%%%%%%%%%%%%%%%%%%%%%%%%%%%%%%%%%%

\begin{abstract}
The structure of the fluid carbon phase in the pressure region of the graphite, diamond, and BC8 solid phase is investigated. We find increasing coordination numbers with an increase in density. From zero to $30\gpa$, the liquid shows a decrease of packing efficiency with increasing temperature. However, for higher pressures, the coordination number increases with increasing temperature.
Up to $1.5\ev$ and independent of the pressure up to $1500\gpa$, a double-peak structure in the ion structure factors exists, indicating persisting covalent bonds. Over the whole pressure range from zero to $3000$ GPa, the fluid structure and properties are strongly determined by such covalent bonds.
\end{abstract}

%\pacs{52.25.Gr,52.70.La}
%%% 52.25.DG - Plasma kinetic equations
%%% 52.25.Kn - Thermodynamics of plasmas
%%% 52.27.Gr 	Strongly-coupled plasmas 
%%% 52.65.Vv 	Perturbative methods
%%% 52.70.La 	X-ray and γ-ray measurements 
%%% 52.25.Mq 	Dielectric properties 

\maketitle

%%%%%%%%%%%%%%%%%%%%%%%%%%%%%%%%%%%%%%%%%%%%%%%%%%%%%%%%%%%%%%%%%%%%%%%%%%%%%%%%
% Introduction
%%%%%%%%%%%%%%%%%%%%%%%%%%%%%%%%%%%%%%%%%%%%%%%%%%%%%%%%%%%%%%%%%%%%%%%%%%%%%%%%
\section{Introduction}
Carbon is the fourth most abundant element in the universe. Hydrocarbons like CH, CH$_2$, or CH$_4$ are therefore important building blocks of the interior of stars and planets \cite{ross:1981,benedetti:1999,hirai:2009,lobanov:2013,Nettelmann:2013,Nettelmann:2016}.
Carbon (as CH) is used as ablation layer in fuel pellets for inertial confinement fusion experiments \cite{lindl:2004}. 
Furthermore, the carbon allotropes graphite and diamond are used in a broad variety of technological processes and experiments, prime example being diamond anvil cells in high-pressure physics \cite{Jayaraman:1983}. 

Therefore, carbon in its many phases and forms has been investigated for a broad range of pressures and temperatures. In particular, the high pressure properties of carbon have attracted a lot of investigations in form of
experiments using laser shocks \cite{alder:1961,bradley:2004,brygoo:2007,hicks:2008,desilva:2009,eggert:2010}, shocks generated using z-pinchs \cite{knudson:2008}, or explosives \cite{kurdyumov:2009,kurdyumov:2012}. These have been combined with VISAR or SOP diagnostics, and, especially in recent years, with x-ray scattering techniques \cite{gregori:2006,pelka:2010,kraus:2012,white:2012,kraus:2013,brown:2014,white:2014,gamboa:2015,Kraus:2016,Kraus:2017,Helfrich:2019}.

Theoretical investigations have been complimentary and are in many ways pioneering as experiments for ultra-high pressures may still be beyond the technical capabilities. We distinguish theories for the equation of state in the high temperature (plasma) regime based on the chemical picture \cite{fried:2000,potekhin:2005,massacrier:2011}, average atom models \cite{thiel:1993,starrett:2013}, density functional molecular dynamics (DFT-MD) for the correlated liquid and solid \cite{galli:1989,galli:1990,scandolo:1995,grumbach:1996,silvestrelli:1998,glosli:1999,wu:2002,wang:2005,correa:2006,romero:2007,correa:2008,knudson:2008,mundy:2008,kraus:2013,benedict:2014,schoettler:2016}, and DFT for a variety of acknowledged and proposed solid phases \cite{fahy:1987,ribeiro:2005,ribeiro:2006,itoh:2009,spanu:2009,zhou:2010,wang:2011,amsler:2012,wang:2012,martinez:2012,zhou:2013,perez:2014}.
More exact theories like quantum Monte Carlo are available \cite{driver:2012,stoyanova:2014} as are classical molecular dynamics simulations employing sophisticated multi-body potentials \cite{morris:1995,ghiringhelli:2005,ghiringhelli:2007,colonna:2009,khaliullin:2011}.

Therefore, there is good understanding of the shock Hugoniot of carbon and its peculiar form following the diamond - fluid phase boundaries near the BC8 phase \cite{alder:1961,bradley:2004,brygoo:2007,hicks:2008,correa:2008,knudson:2008,romero:2007,desilva:2009,eggert:2010}. The reflectivity \cite{bradley:2004} and conductivity \cite{desilva:2009} have been measured
along the Hugoniot and in off-Hugoniot states \cite{reitze:1992,haun:2002} connecting changes in structure with increases in the reflectivity and conductivity. DFT calculations and XRTS measurements show that the band gap in solid diamond increases with pressure but decreases with temperature to the point that the band gap is closed at melting \cite{correa:2006,romero:2007,gamboa:2015,ramakrishna:2019}. 

Particular attention has been given to the determination of the melt line of carbon from ambient pressure to several  megabars \cite{grumbach:1996,fried:2000,ghiringhelli:2005,wang:2005,sava:2005,correa:2006,brygoo:2007,hicks:2008,correa:2008,colonna:2009,eggert:2010,benedict:2014,Helfrich:2019}. Quantitative details differ from theory to theory but it is clear that graphite as well as diamond and several other high pressure phases of carbon exhibit a melting line maximum. The exact position of the diamond-BC8-liquid triple point is debated \cite{knudson:2008} as is the position, but not the existence, of the liquid-liquid phase transition in the range of $1-5\gpa$ \cite{thiel:1993,glosli:1999,wu:2002}. Recently, there has been a report on a further monomer-dimer transition around $7\ev$ in the low density carbon plasma \cite{dharma:2017}.

The graphite phase \cite{spanu:2009}, the graphite to diamond transition \cite{scandolo:1995,mundy:2008,wang:2011,winey:2013}, the diamond phase \cite{erskine:1991,ribeiro:2005,ribeiro:2006,kurdyumov:2009,kurdyumov:2012,martinez:2012,stoyanova:2014} and further high pressure and exotic phases have been studied \cite{erskine:1991,ribeiro:2005,ribeiro:2006,kurdyumov:2009,zhou:2010,kurdyumov:2012,martinez:2012,wang:2012}. The phase transition properties under quasi-static condition have been compared with ultra-fast processes and nucleation mechanisms in order to explain and expand on the first \cite{silvestrelli:1998,ghiringhelli:2007,mundy:2008,pelka:2010,khaliullin:2011,hauriege:2012}.
 
As stated before, there is a phase transition in the low density fluid with a critical temperature between $5500\,$K and $9000\,$K \cite{thiel:1993,glosli:1999,wu:2002}. The structure of the fluid has been characterized over a broader pressure range showing an increase in coordination from graphite-like (3-fold coordination) to diamond-like (4-fold coordination) and beyond \cite{grumbach:1996,galli:1989,galli:1990,morris:1995}. The anisotropic nature of the low density fluid has been pointed out in the bond angle distribution \cite{galli:1989}. There has only very recently been a direct measurement using XRTS of the structure of the fluid around $1\Mbar$ hinting at a substantially more structured/complicated fluid than expected \cite{kraus:2012,kraus:2013}.

Short-pulse, high-power lasers coupled with high quality x-ray scattering enable nowadays also the study of structural and temperature relaxation in carbon as direct precursor for equilibrium processes \cite{white:2012,brown:2014,white:2014}. Similarly, strain and strength investigations of diamond provide new insight for materials research \cite{tsay:1977,wang:2010,zhou:2013,perez:2014,gamboa:2015,macdonald:2016}.

Especially in light of new x-ray scattering capabilities using x-ray free electrons lasers (XFEL), we investigate the structure of the carbon fluid in the parameter range from $10$ to $3000\gpa$ and temperatures from $0.8$ to $5\ev$ by performing large-scale DFT-MD simulations. In particular, we analyse the effect of the large number of temporary covalent bonds on the ion-ion structure factor. We examine the clustering in the fluid and especially study the electron-ion distributions and the electron localization function. Further, we calculate the electronic DOS and the conductivity. In this way we reveal details of the compression behavior of dense fluid carbon near the high-pressure melting line.

%\jan{
%ToDo: triple point graph-diamond-liq \cite{prawer:1992}
%occurence in planets (CH \cite{benedetti:1999,hirai:2009,lobanov:2013}, 
%diamond rain \cite{ross:1981}, Nadine's papers on Uranus and Neptune \cite{Nettelmann:2013,Nettelmann:2016})\\
%ToDo: Aim and outline of the paper
%}
%%%%%%%%%%%%%%%%%%%%%%%%%%%%%%%%%%%%%%%%%%%%%%%%%%%%%%%%%%%%%%%%%%%%%%%%%%%%%%%%
% methods
%%%%%%%%%%%%%%%%%%%%%%%%%%%%%%%%%%%%%%%%%%%%%%%%%%%%%%%%%%%%%%%%%%%%%%%%%%%%%%%%
\section{Methods}
For our calculations, we use the finite temperature DFT-MD (FT-DFT-MD) framework which combines classical molecular dynamics for the ions with a quantum treatment of the electrons based on FT-DFT. 
We use its implementation in the Vienna \textit{ab initio} simulation package VASP~5.3.~\cite{Kresse:1993, Kresse:1994, Kresse:1996a} and the provided projector augmented wave~\cite{Blochl:1994, Kresse:1999} pseudopotential for the interaction between the nuclei and the electrons. The exchange and correlation term is approximated by the generalized gradient functional of Perdew, Burke and Ernzerhof~\cite{Perdew:1996} for all simulations.
The simulations were carried out using 192 to 256 atoms, considering 4 valence electrons per atom, an energy cutoff of $1100\ev$ and a simulation time of $1$ to $5\,$ps. The ion temperature was controlled with a Nos\'{e} thermostat~\cite{Nose:1984}. Particular care was taken to include enough empty bands for computations at elevated temperatures. Evaluations of the Brillouin zone were performed at the Baldereschi mean value point~\cite{Baldereschi:1973}. The simulations were performed along isochores for densities of $2.5$~g/cm$^3$, $3$~g/cm$^3$, $4$~g/cm$^3$, $5$~g/cm$^3$, $6$~g/cm$^3$, $7$~g/cm$^3$, $8$~g/cm$^3$, and $10$~g/cm$^3$ with temperatures $ T $ rising from $0.9\ev$ to $4\ev$. 
\newline
% % % % % % % % % % % % % % % % % % % % % % % % 
% Pair distribution
% % % % % % % % % % % % % % % % % % % % % % % % 
With \textit{ab initio} FT-DFT-MD simulations, the pair correlation function of the ions 
$ g_\mathrm{ii}(r) $ is 
accessible by averaging over all ions and simulation steps in equilibrium,
\begin{equation}\label{eq:paar_ionen}
	g_\mathrm{ii}(r) = \frac{ V } { 4 \pi r^2 N_\mathrm{i} ( N_\mathrm{i} - 1 ) }  \left<  \sum^{N_\mathrm{i}}_{a=1} \sum^{N_\mathrm{i}}_{b = 1 \atop b \neq a}  \delta( \vec{r} - \vec{R}_{ab} ) \right>	\,,
\end{equation}
where $N_\mathrm{i}$ denotes the particle number and V the cell volume and $ \vec{R}_{ab} = | \vec{R}_a - \vec{R}_b | $. 

From the pair correlation function, the coordination number may be obtained via
\beq
N_C=4\pi n_i \int\limits_0^{R_c} r^2 dr\, g(r)\,.
\eeq
Here, $n_i$ is the ion number density and the upper limit of integration is given by the position of the first minimum of the pair correlation function.

The respective static ion-ion structure factor $S_{ii}(k)$ were obtained via the Fourier transform of the ion-ion pair correlation function,
\begin{equation}\label{Struc}
	S_{\mathrm{ii}}(k) = 1 + n_\mathrm{i} \int g_{\mathrm{ii}}(r) e^{i\vec{k}\vec{r}}\mathrm d\vec{r}	\,.
\end{equation}

% % % % % % % % % % % % % % % % % % % % % % % % 
% coherent sets
% % % % % % % % % % % % % % % % % % % % % % % % 
\begin{figure}[th]
\begin{center}
\includegraphics[width=0.95\columnwidth]{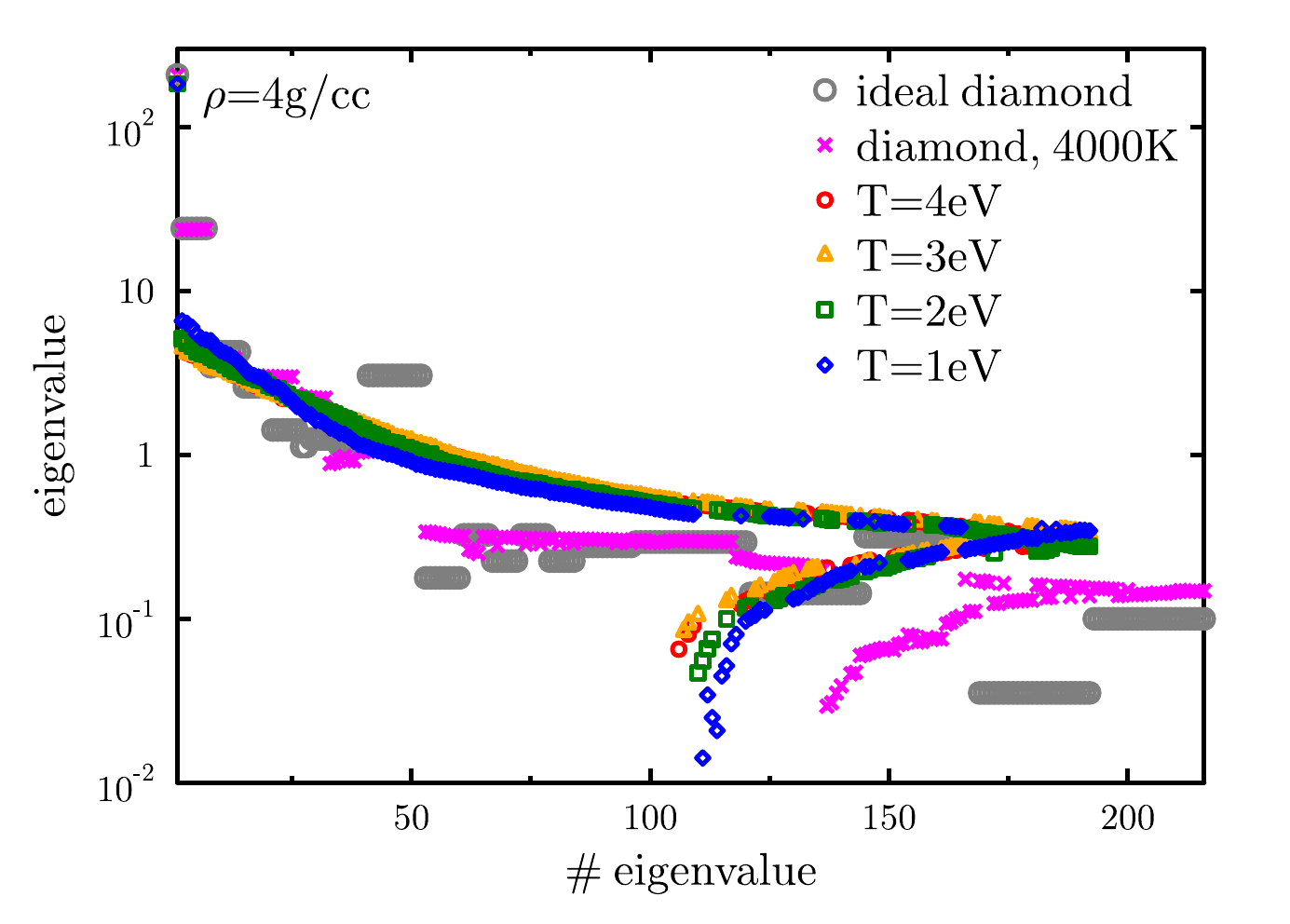}
\caption{The eigenvalues of the distance matrix according to coherent set analysis for an ideal diamond lattice, for diamond at $T=4000$ K, and for the dense carbon fluid, all at $4$ g/cm.}
\label{fig:coh_ev}
\end{center}
\end{figure}
A cluster analysis can reveal further details about the ion structure and the bonds forming between carbon atoms. Instead of using a cluster analysis based on a (free parameter) distance cut-off, we employ the method of coherent sets \cite{Froyland:2013,Denner:2016,Denner:2017,
Fackeldey:2017}. After building the matrix of all pairwise interparticle distances (either in location space or in momentum/velocity space), the eigenvalues and eigenvectors of this matrix are determined. The resulting eigenvalues show a characteristic signature indicating the cluster structure, see Fig.\ref{fig:coh_ev}. Lattices show a specific signature with degenerate eigenvalues, an ideal gas would have N identical eigenvalues of unity. The eigenvectors can be used to identify and plot clusters (coherent sets) within the DFT-MD supercell. The range of values of the product of a particular eigenvalue and it's eigenvector can be divided into bins and thus particles belonging to the same cluster will be identified. We chose to divide the range of the second eigenvector into 20 bins, the qualitative results do not depend on the number of bins.

% 
% % % % % % % % % % % % % % % % % % % % % % % % 
% electron ion structure
% % % % % % % % % % % % % % % % % % % % % % % % 
Similarly to the ion structure, the structure of the electrons in relation to the ions is of relevance. We determine the electron-ion pair correlation function via
\begin{equation}\label{eq:paar_ei}
	g_\mathrm{ei}(r) = \frac{ V } { 4 \pi r^2 N_\mathrm{e} ( N_\mathrm{i} - 1 ) }  
	\int d\vec{r}\,'\left< n_e(\vec{r}-\vec{r}\,')n_i(\vec{r}\,')  \right>	\,.
\end{equation}

% % % % % % % % % % % % % % % % % % % % % % % % 
% ELF
% % % % % % % % % % % % % % % % % % % % % % % % 
We also use the electron localization function to investigate the bonding structure in carbon at high pressure \cite{Becke:1990}.
% % % % % % % % % % % % % % % % % % % % % % % % 
% DOS
% % % % % % % % % % % % % % % % % % % % % % % % 
The electronic density of states (DOS) shows the distribution of eigenstates in energy space and was determined by summing all the histograms of eigenvalues of all MD snapshots for a particular density and temperature.

% % % % % % % % % % % % % % % % % % % % % % % % 
% Conductivity
% % % % % % % % % % % % % % % % % % % % % % % % 
In an additional analysis, we calculated the electrical conductivity. The electronic conductivity is directly calculated from FT-DFT eigenvalues and wave functions via the Kubo-Greenwood formula~\cite{Kubo:1957,Greenwood:1958},
\begin{align}\label{eq:Kubo-Green}
	\sigma_\mathrm{e}(\omega) = &\frac{ 2 \pi } { 3 m_\mathrm{e} V \omega } \sum_{\vec{k}} w_{\vec{k}} \sum_\nu^{N_b} \sum_\mu^{N_b} ( f_{\vec{k} \nu} - f_{\vec{k} \mu} ) 	\nonumber	\\ 
&\times	\langle \phi_{ \vec{k} \nu } | \hat{\vec{p}} | \phi_{ \vec{k} \mu } \rangle^2 \delta \left( \epsilon_{ \vec{k} _\mu} - \epsilon_{ \vec{k} _\nu} - \hbar \omega \right)	\,,
\end{align}
where $m_\mathrm{e}$ is the electronic mass, $V$ the volume of the cubic simulation cell and $\omega$ the frequency. The indices $\nu$ and $\mu$ run over all $N_b$ bands and sum the matrix elements of the Bloch functions with the momentum operator weighted by the difference of the Fermi occupations $ f_{\vec{k} \mu} $ of the bands. The integration over the first Brillouin zone is evaluated with a discrete mesh of $\vec{k}$ points and their respective weights $ w_{\vec{k}} $. Since all bands have discrete eigenvalues, the $ \delta $ function hast to be broadened 
to a finite width. For each different temperature and density point we applied Eq. (\ref{eq:Kubo-Green}) on the FT-DFT simulations from 10-20 different ion configurations. The respective results are then averaged. The ion configurations were taken from converged simulation runs in thermodynamic equilibrium. 

% % % % % % % % % % % % % % % % % % % % % % % % 
% Diffusion & viscosity
% % % % % % % % % % % % % % % % % % % % % % % % 
%We use Green-Kubo relations to calculate self diffusion and viscosity. The velocity autocorrelation function provides the self-diffusion coefficient
%\beq
%D=\frac{1}{3 N_i}\int\limits_3^{\infty} dt\, \langle {\bf v}_i(0){\bf v}_i(t)\rangle
%\eeq
%with the ensemble average $\langle\ldots\rangle$ over velocities ${\bf v}_i$.
%The viscosity tensor can be obtained similarly from
%\beq
%\eta=\frac{V}{3 k_BT}\int\limits_3^{\infty} dt\, \Sigma_{\alpha\beta} 
%\langle \sigma_{\alpha\beta}(0)\sigma_{\alpha\beta}(t)\rangle\,,
%\eeq
%where $\sigma$ is the stress tensor and with the off-diagonal elements $\alpha\beta=\{xy,xz,yx,yz,zx,zy\}$.
%%%%%%%%%%%%%%%%%%%%%%%%%%%%%%%%%%%%%%%%%%%%%%%%%%%%%%%%%%%%%%%%%%%%%%%%%%%%%%%%
% results
%%%%%%%%%%%%%%%%%%%%%%%%%%%%%%%%%%%%%%%%%%%%%%%%%%%%%%%%%%%%%%%%%%%%%%%%%%%%%%%%
\section{Results}
\begin{figure}[th]
\begin{center}
\includegraphics[width=0.95\columnwidth]{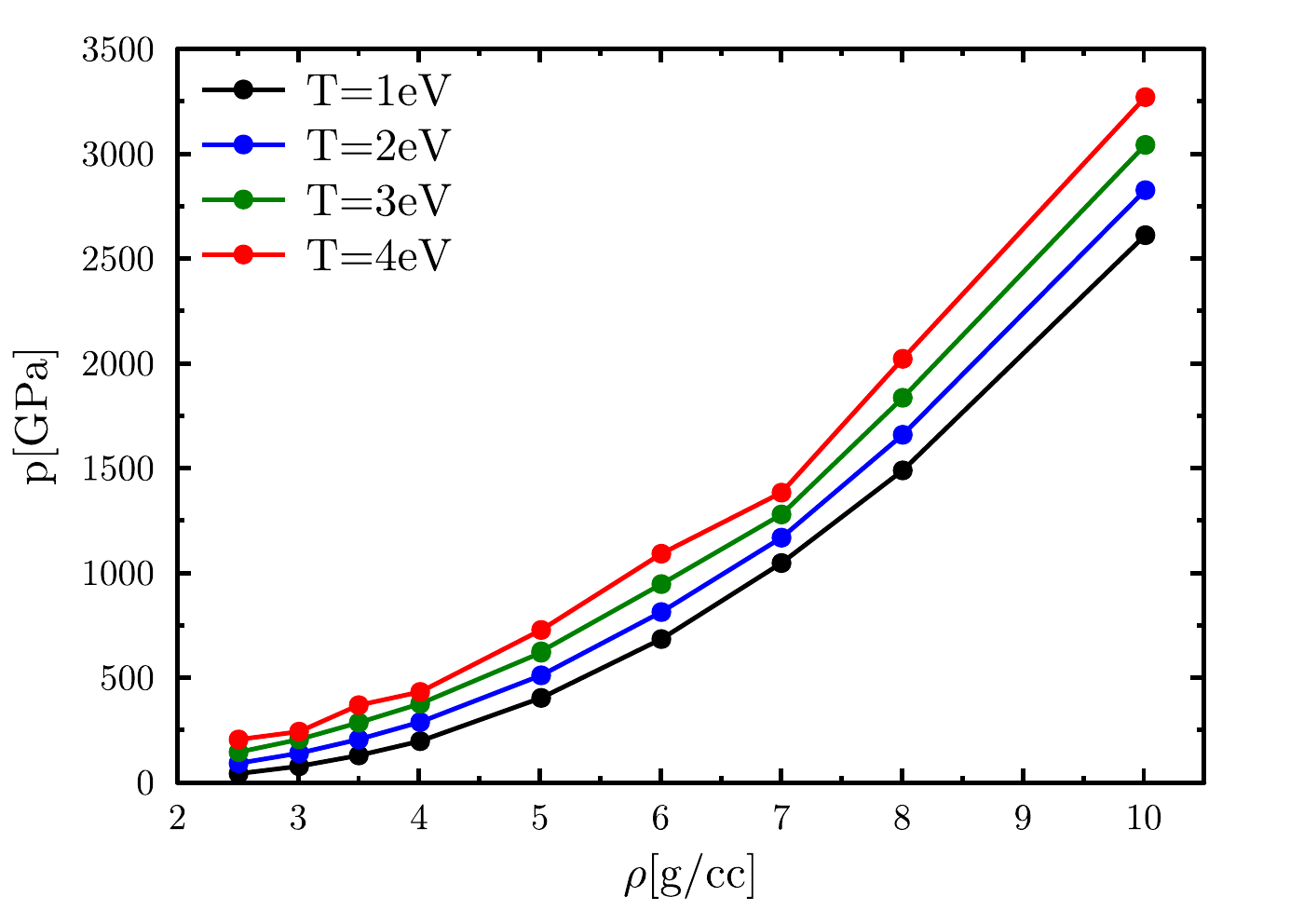}
\caption{Four different isotherms of the carbon fluid in the Mbar range.}
\label{fig:eos}
\end{center}
\end{figure}

We analyse the ionic structure of fluid carbon in position and momentum space. The emerging picture is that of a very complex fluid at high pressure, dominated by temporary covalent bonds and transient clusters. In Fig. \ref{fig:eos}, the density, temperature, and pressure range considered is shown. It spans the fluid region at high temperature corresponding to the graphite, diamond, and BC8 solid phases.
%\jan{say something about good agreement with EOS of Benedict et al.}

\subsection{Pair correlation function}
\begin{figure*}[th]
\begin{center}
\includegraphics[width=0.48\columnwidth]{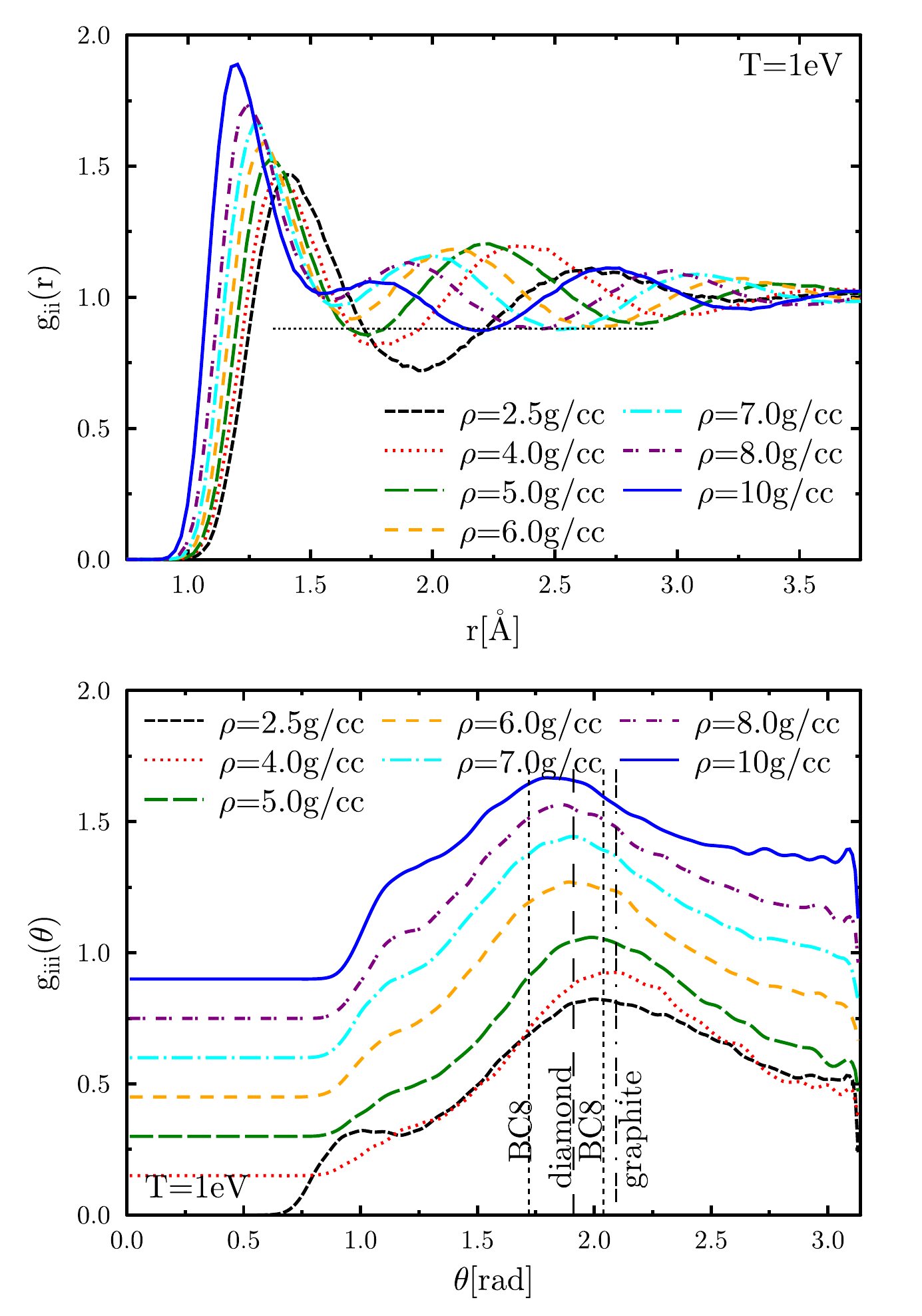}
\includegraphics[width=0.48\columnwidth]{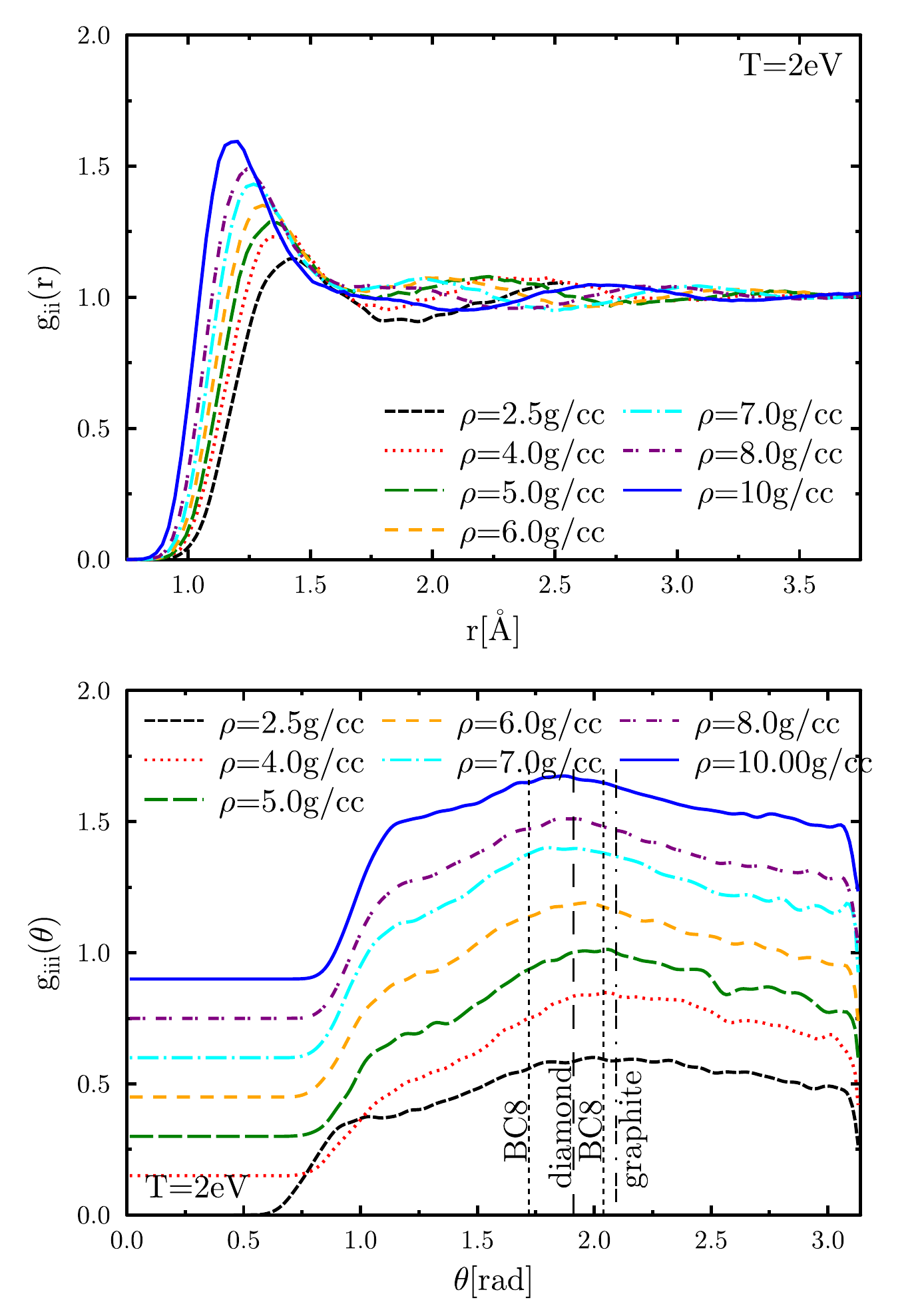}
\caption{Pair correlation functions and bond angle distribution in the liquid for two different temperatures as function of density. The horizontal dotted line in the top left panel is a guide to the eye. The bond angle distributions for different densities are shifted along the y-axis for better visibility. Bond angles occurring in the solid phases are indicated by vertical dashed, dotted, and dash-dotted lines, respectively.}
\label{fig:giii}
\end{center}
\end{figure*}
Fig. \ref{fig:giii} shows the pair correlation functions and bond angle distributions of fluid carbon for a wide range of parameters from $p=30\gpa$ to $p=3000\gpa$ and up to a temperature of $T=2\ev$.
At $T=1\ev$ and below, from $2.5\gcc$ to $4\gcc$, the height of the first peak of the pair correlation function does not increase. This is the fluid similar to the graphite solid phase in which any density increase is given by a reduction in the interplanar distance of graphene sheets. For higher densities, the height of the first $g(r)$ peak increases as all neighbors are pushed closer to each other. 
An increase in temperature decreases the short-range order in the fluid and therefore reduces the peaks and valleys in the pair correlation.

\begin{figure*}[t]
\begin{center}
% reprint
%\includegraphics[width=0.95\columnwidth]{first_neighbor_3.pdf}\\
%\includegraphics[width=0.95\columnwidth]{pt_coord_surf_2.pdf}
\includegraphics[width=0.48\columnwidth]{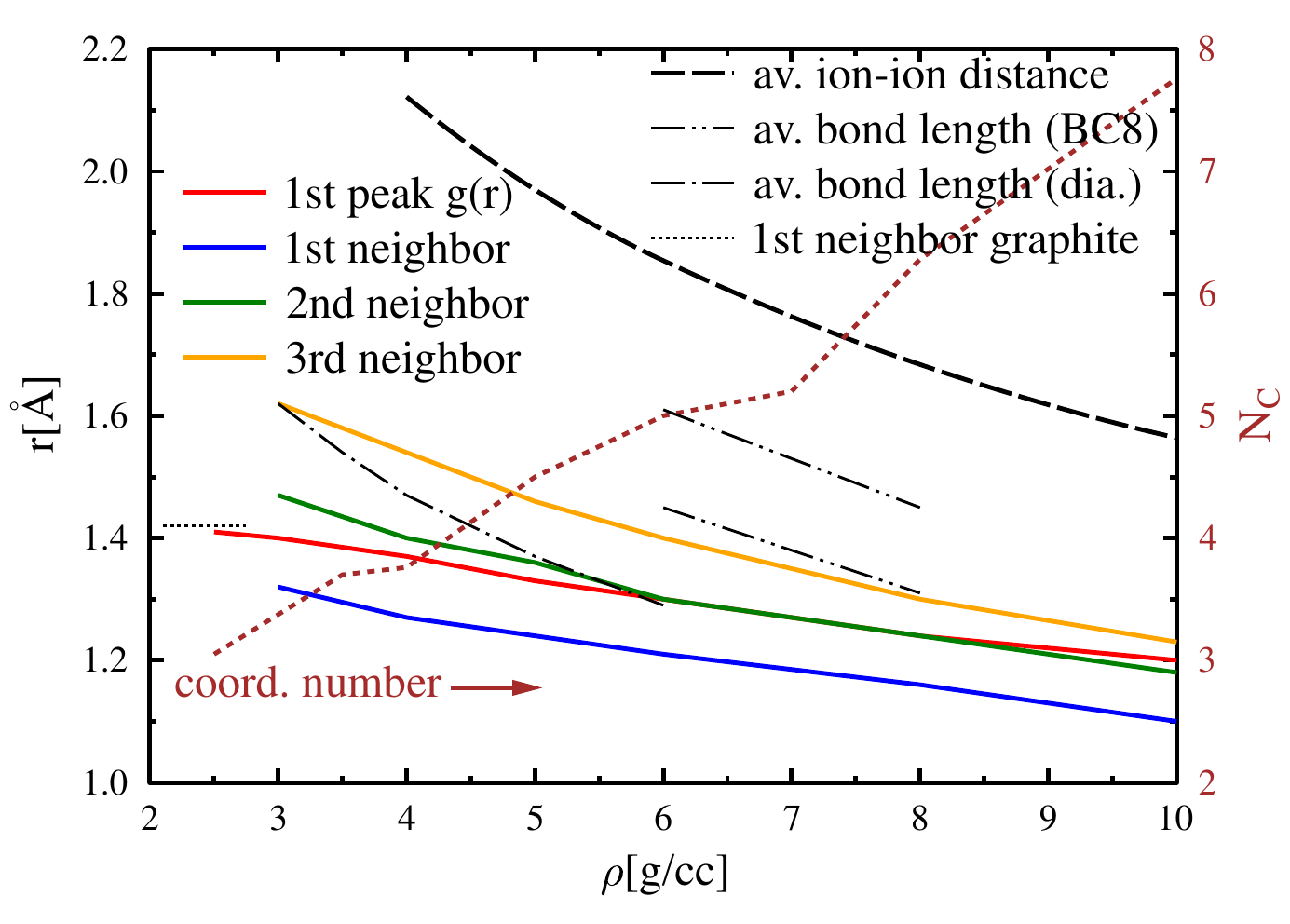}
\includegraphics[width=0.48\columnwidth]{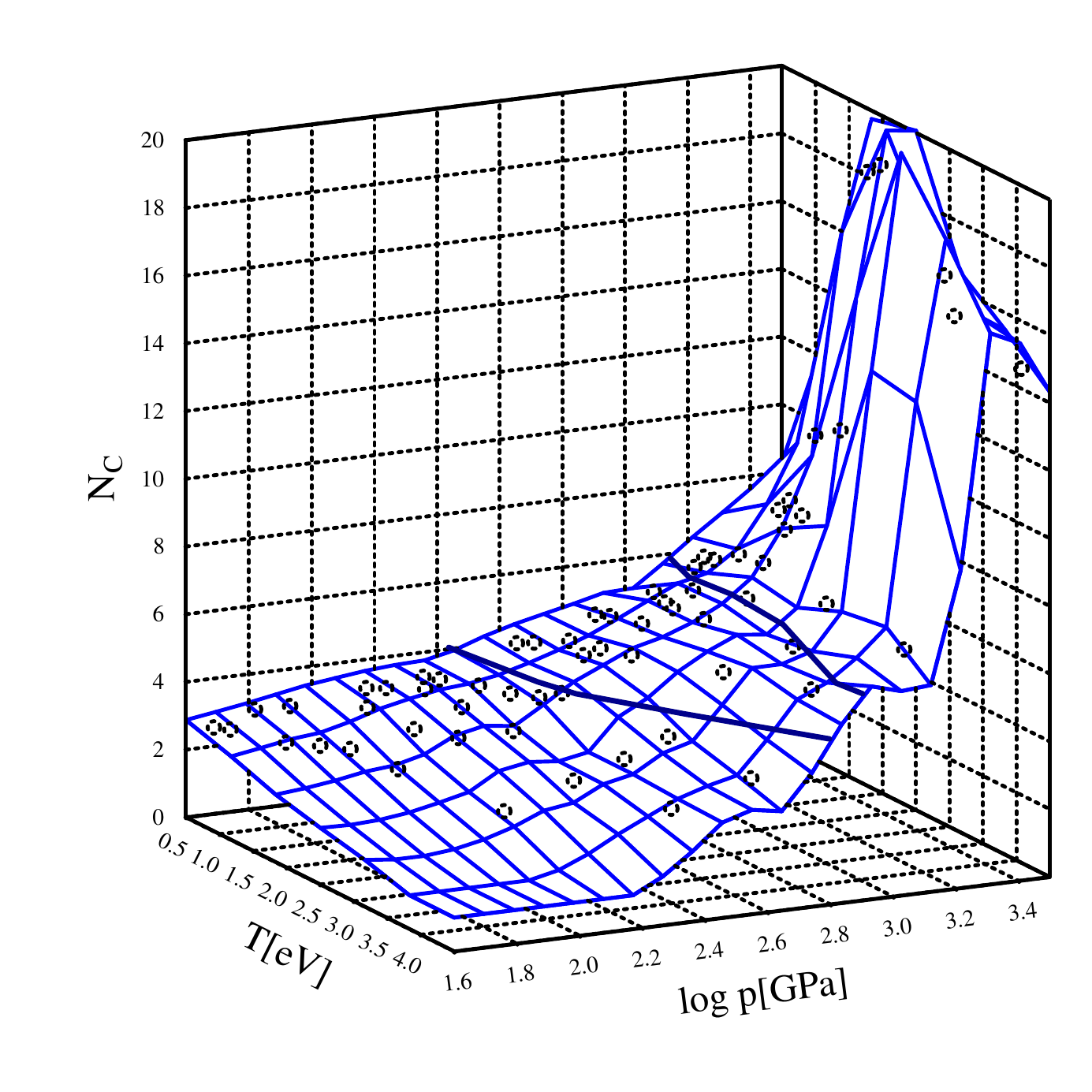}
\caption{(Top) Location of 1st, 2nd, and 3rd neighbor shell at $T=1\ev$ in relation to the peak of the pair correlation function, the bond length, the average ion-ion distance $d=(3/4\pi n_i)^{1/3}$, and the coordination number $N_C$. (Bottom) The coordination number as function of temperature and pressure.}
\label{fig:first_neighbor}
\end{center}
\end{figure*}

The first peak in the pair correlation function (see also Fig. \ref{fig:first_neighbor}) changes its location from $2.5\gcc$ to $10\gcc$ only slightly from $1.4\,\AA$ (close to the zero pressure graphite bond length of $1.42\,\AA$) to $1.2\,\AA$. It does so smoothly and without the abrupt changes bond lengths undergo during phase changes from one solid phase to the next. If the first peak had followed the reduction in the spatial dimension due to density increase (as would be the case in an atomic or ionic fluid {\em not} dominated by bonds), it should have shrunk to $0.87\,\AA$. This shows that the first peak in the pair correlation function is given by the carbon bond and the resistance of carbon bonds to pressure in general. The breaking of carbon bonds due to the pressure is negligible, even for $10\gcc$ and $2.6\Mbar$.

The temperature influence on the location of the first peak of the pair correlation is very small over the entire density range considered here. Only the distance to the first neighbor becomes significantly smaller with increasing temperature up to $4\ev$. For $\rho=6\gcc$, we have analysed structural data up to a temperature of $T=6\ev$, and even for this temperature, the position of the first maximum does not change. As this position is given by the bond length, and the first peak in a partially ionized system dominated by screened Coulomb interactions would be at considerably larger distances, we again have to conclude the absence of free electrons in significant numbers up to at least $T=6\ev$.

Apparent is further the non-standard form of the pair correlation function for distances larger than the first maximum. In particular for densities larger than 
$\rho=5\gcc$, there is an asymmetry in the flanks of the first peak. Further, the first minimum following the first peak is only a local minimum and the global minimum is located after the second peak of the pair correlation function (see horizontal line in the top left panel of Fig. \ref{fig:giii} as guide to the eye). This is structural behavior contrary to the carbon fluid at smaller densities and to fluids  whose structures are dominated by Coulomb or soft-core potentials, i.e. plasmas, metallic fluids, or neutral fluids. In the latter examples, the maxima and minima would follow a damped oscillation.

In addition to the order in distance as apparent in the pair correlation function, there is angular ordering in the bonds as can be seen from the bond angle distributions. The bond angles within a triplet of carbon atoms in the fluid follow the bond angles that occur in the different solid phases at similar densities. This means that there is a shift of the maximum in the bond angle distributions from around $120^{\circ}$ (graphite, $2.09$ rad) via $107^{\circ}$ (diamond, $1.87$ rad) to $98.5^{\circ}$ (BC8, $1.72$ rad). Again, temperature widens the peak in the bond angle distribution and allows more bond angles with similar probability. The sharp drop around $60^{\circ}\hat{=} 1.05$ occurs when all bond lengths in the triangle formed by three carbon atoms are the same. Smaller bond angles (particularly smaller than $\approx 40^{\circ}$) are discouraged by Pauli repulsion of the full $1s^2$ orbitals. The minimum angle is consistent with the approximate extension of the $1s$-orbitals of $1 a_B$. 
The left flank of the bond angle distribution steepens with increasing density and slightly increasing the minimum allowed angle reflecting the effects of stronger correlations for higher densities.
%In addition, as the bond length becomes smaller with increasing density, the minimum angle increases slightly. This is consistent with a constant extension of the $1s$ states which remain stable up to the highest densities and temperatures considered here.

Such carbon-bond dominated structure displays some interesting length scales as shown in Fig. \ref{fig:first_neighbor}. The distances to the first, second, and third neighbor as well as the first peak of the pair correlation function are all substantially smaller than the average ion-ion distance calculated from the mean density. The ratio of the distance of the first peak of $g(r)$ to the average ion-ion distance is $f=0.77$ for carbon at a density of $\rho=10\gcc$ and falls to $f=0.56$ for a density of $\rho=3\gcc$ at a temperature of $T=1\ev$. In comparison, this ratio in warm dense (metallic) hydrogen is $f=0.81$ \cite{Vorberger:2013}, in warm dense lithium $f=0.81-0.86$ \cite{kietzmann:2008}, warm dense aluminium $f=0.8$ \cite{fletcher:2015}, warm dense iron $f=0.8-0.86$ \cite{white:2017}, warm dense silicon $f=0.83$ \footnote{own calculations, unpublished}.
Keeping in mind the even shorter distance to the first neighbor in the fluid, this again points to the dominating influence of the carbon-carbon bonds.

\begin{figure*}[t]
\begin{center}
\includegraphics[width=0.48\columnwidth]{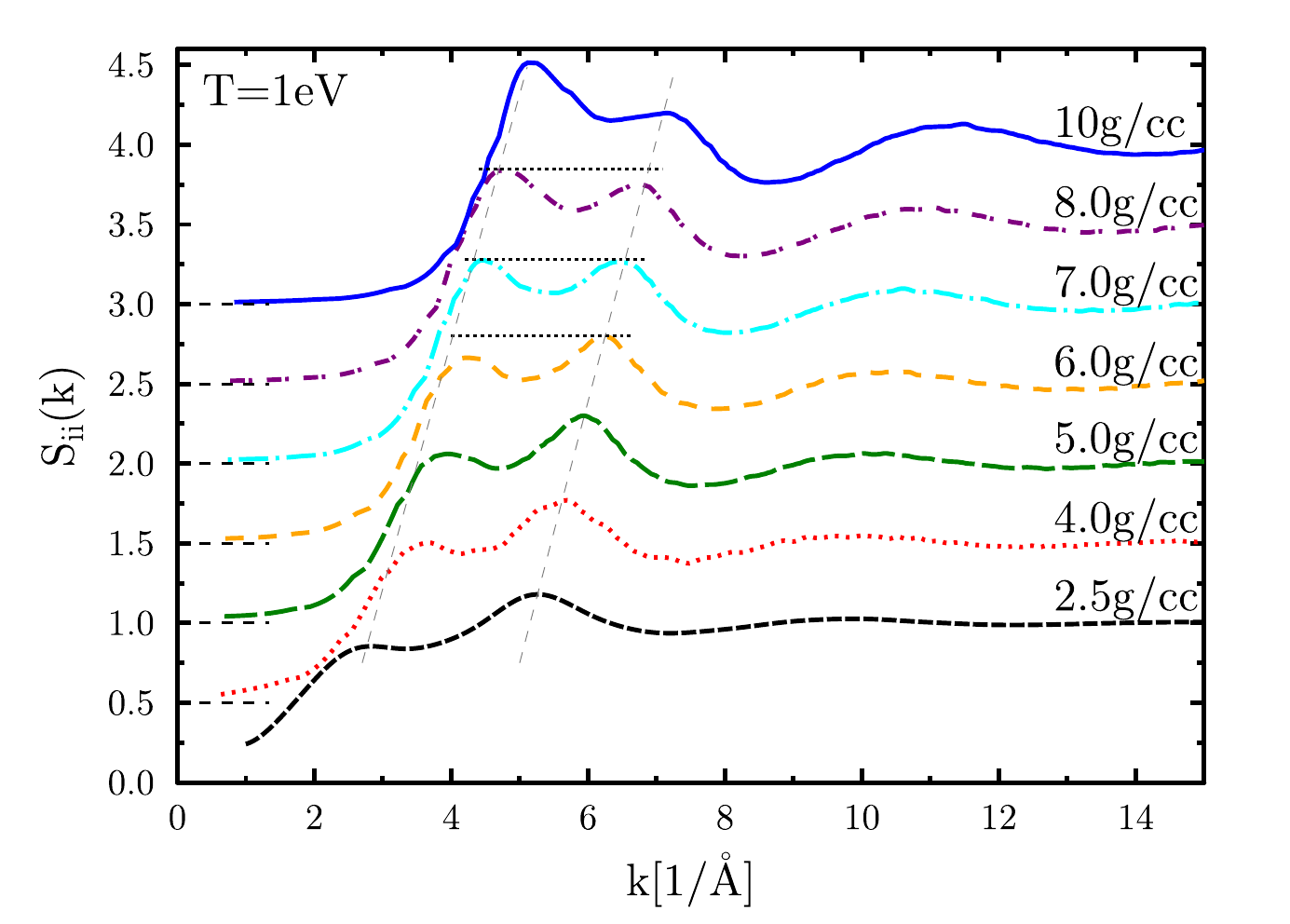}
\includegraphics[width=0.48\columnwidth]{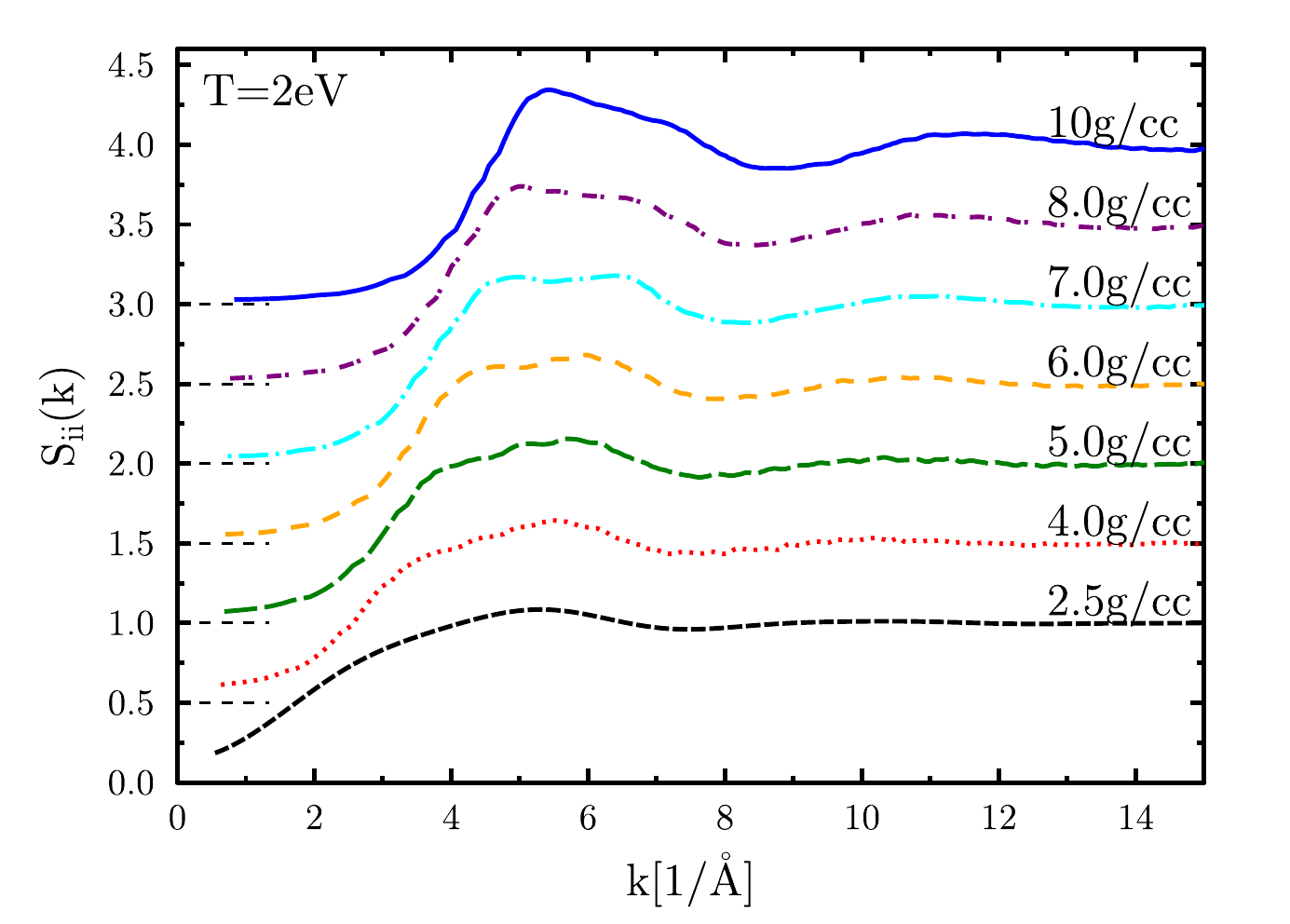}
\includegraphics[width=0.48\columnwidth]{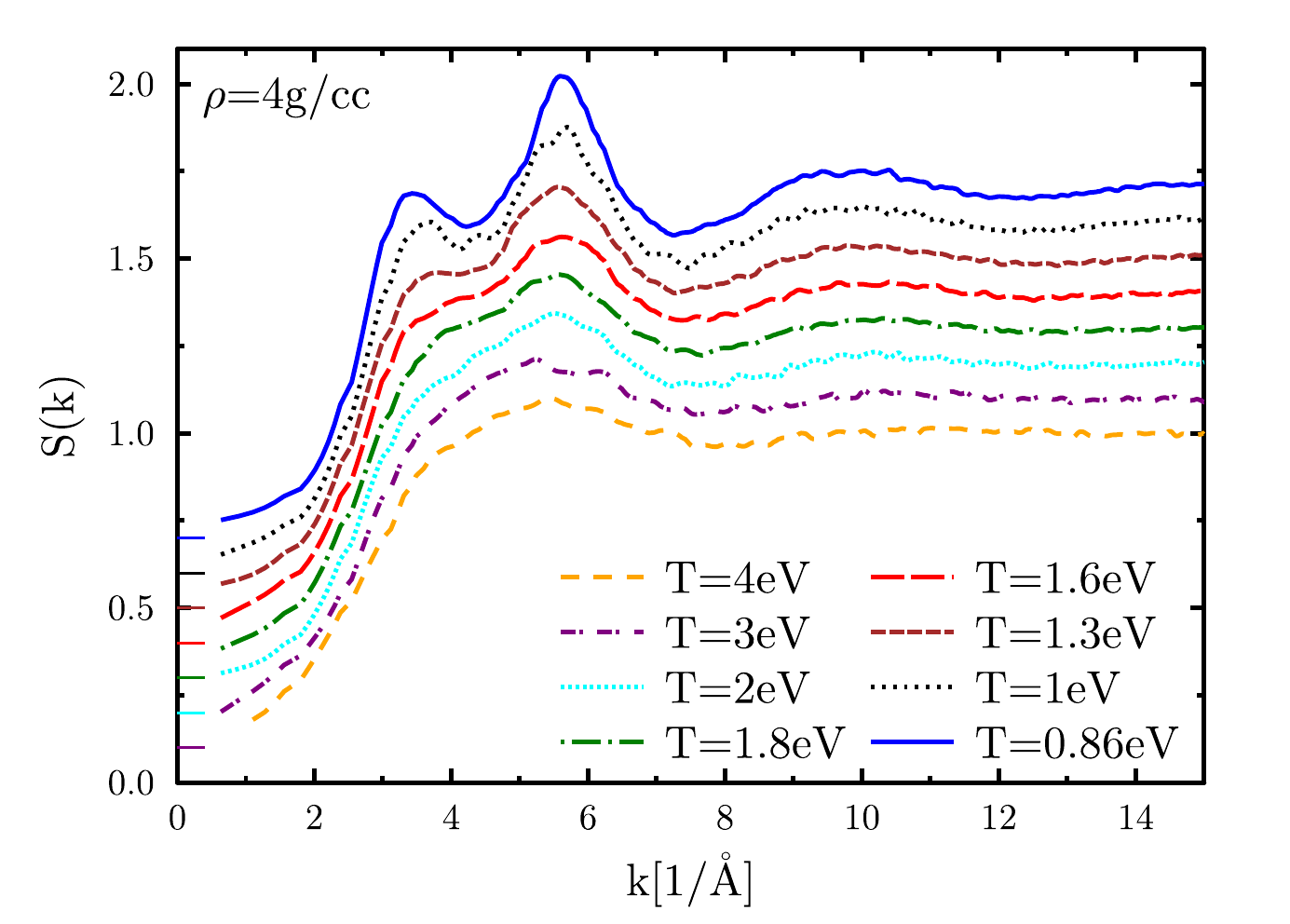}
\includegraphics[width=0.48\columnwidth]{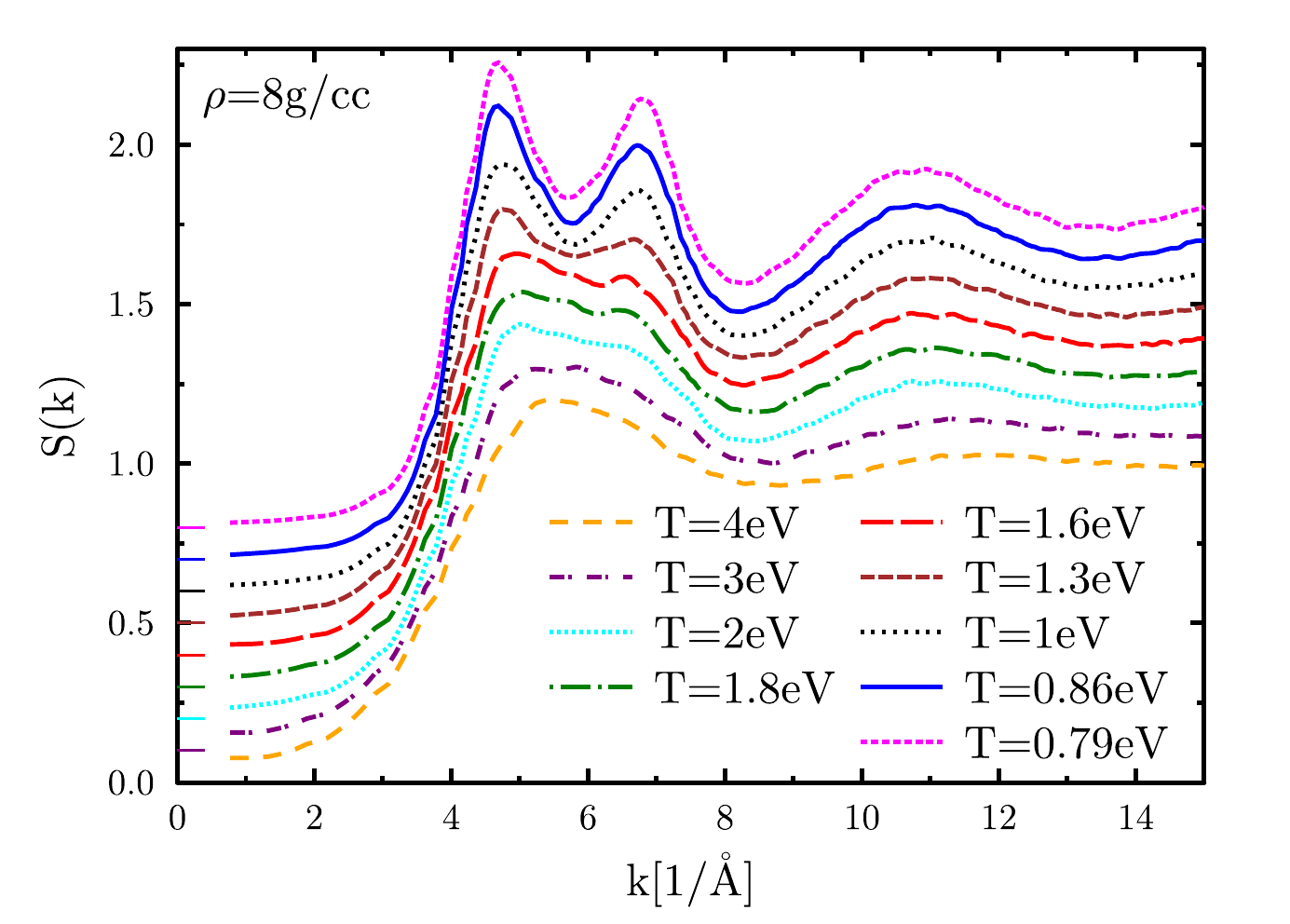}
\caption{(Top Row) Static structure factors for fluid carbon for two different temperatures as function of density. (Bottom Row) Static structure for two different densities as function of temperature. Each structure factor is shifted along the y-axis. The zero line for each structure factor is indicated by a short horizontal line on the left.}
\label{fig:sii}
\end{center}
\end{figure*}
The non-standard short-range order also leads to some non-standard coordination numbers, see the bottom panel of Fig. \ref{fig:first_neighbor}. For the low pressure fluid, a graphite-like coordination number is observed and the coordination numbers drop as the temperature is raised. However, this changes around a pressure of $p=3\Mbar$ and the coordination number in the liquid is  higher than in the diamond or BC8 solid phases for pressures to $p=20\Mbar$ and beyond. In addition, the temperature dependence of the coordination number is unusual in the range up to $T=4\ev$ since temperature has almost no effect; if at all, a temperature increase leads to an increased coordination number (see also the discussion of the carbon phase space below).
\subsection{Static structure factor}
This property can be better understood while analysing the static structure factors in Fig. \ref{fig:sii}. Here, the mid- and long-range order appear in the structure factors as a temperature dependent double peak structure. The double peaks are remnants of Bragg peaks of the graphite, diamond and BC8 phases in the solid. Thus, a temporary short- and mid-range order persists in the liquid that is similar to the solid phases at the appropriate pressure. 

For small temperatures in the liquid, for which such a double peak structure can generally be observed, it is apparent that the second peak represents the global maximum for densities lower than $\rho=7\gcc$. At $\rho=7\gcc$, both peaks are equal in height and the first peak dominates for densities larger than $\rho=7\gcc$. This can be put into relation with the relative importance of the first few Bragg peaks in the diamond and BC8 phases, respectively. The solid equivalent of the first fluid peak from $\rho=4\gcc$ to $\rho=7\gcc$ is the $(111)$ Bragg peak in diamond and the second fluid peak is the peak of the $(220)$ Bragg reflection in diamond. From approximately $p=10\Mbar$ the fluid's structure is BC8 like with the remnants of the $(211)$, $(222)$ \& $(312)$ Bragg peaks dominating the fluid structure and making the first peak the global maximum.

The lowest density for which a structure factor is displayed in Fig. \ref{fig:sii}, namely $\rho=2.5\gcc$, is lower than that for solid diamond in equilibrium. The fluid structure factor however very much resembles the diamond-like fluid with a dominating second peak. The structure can best be described as the fluid analogue to the hexagonal diamond phase with a strongly reduced c-distance as compared to graphite.

In the temperature range from $T=1.6\ev$ to $T=2\ev$, the double peak structure can be observed to vanish over the whole pressure range considered here. Within the uncertainty of the simulations, there is no pressure or density dependence in this process. Similarly to what was concluded while analysing the pair correlation functions, this shows the resilience of the carbon-carbon bonds to pressure and the dominating role of the bonds for the liquid in the entire pressure range that was considered here.

For temperatures where a double peak structure does not appear, the first peak in the static structure is broadened to a great degree. Even at temperatures of $T=4\ev$ and above, the structure factor's first peak remains asymmetrical with a steep first slope and a broad right flank. 

\subsection{Transient carbon clusters}
\begin{figure*}[t]
\begin{center}
% reprint
%\includegraphics[width=\textwidth]{coh_set_cluster_bin.pdf}
\includegraphics[width=\textwidth]{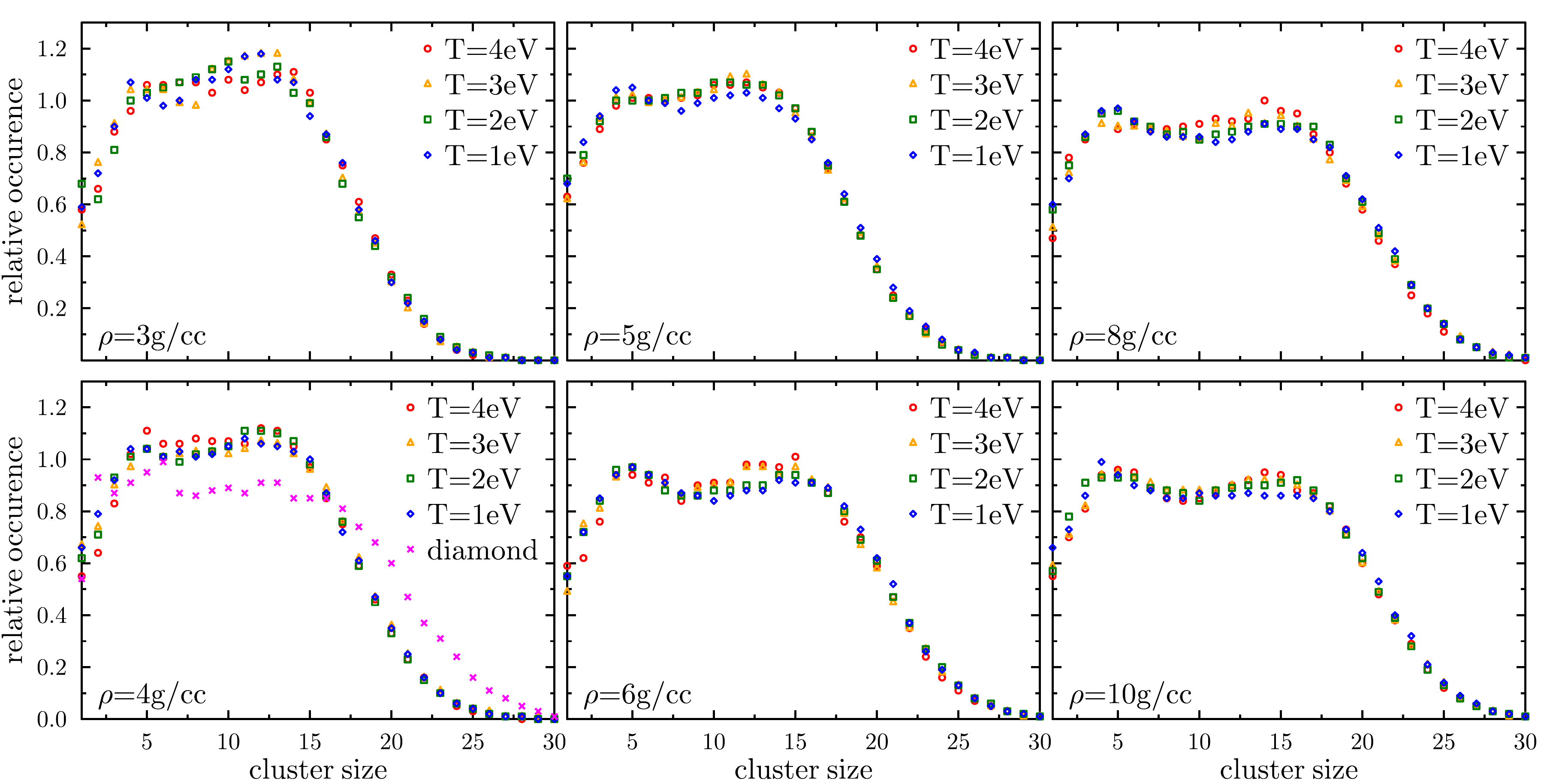}
\caption{Relative occurrences of clusters of different sizes according to the analysis of coherent sets for densities between $3$ g/cc and $10$ g/cc and temperatures from $1$ eV to $4$ eV.}
\label{fig:cluster}
\end{center}
\end{figure*}
As carbon atoms can establish a variety of different types of bonds to a number of nearest neighbors, the fluid might also be understood as a mixture of transient carbon clusters with varying number of atoms. The method of coherent sets delivers the clustering within the supercell without the need of free parameters and the results are shown in Fig. \ref{fig:cluster}.

The comparison of a finite temperature solid diamond system with the fluid carbon shows that smaller clusters are prevalent in the fluid. We also find a broadening of the cluster size distribution with density such that a double peak distribution arises. For lower densities, the maximum is located at around $N=13$, whereas for the highest densities, it is at $N=3$. Interestingly, the temperature dependence of the cluster distributions is negligible.
\subsection{Electron-ion correlations}
\begin{figure*}[t]
\begin{center}
% reprint
%\includegraphics[width=\textwidth]{gei_rho_2.pdf}
\includegraphics[width=\textwidth]{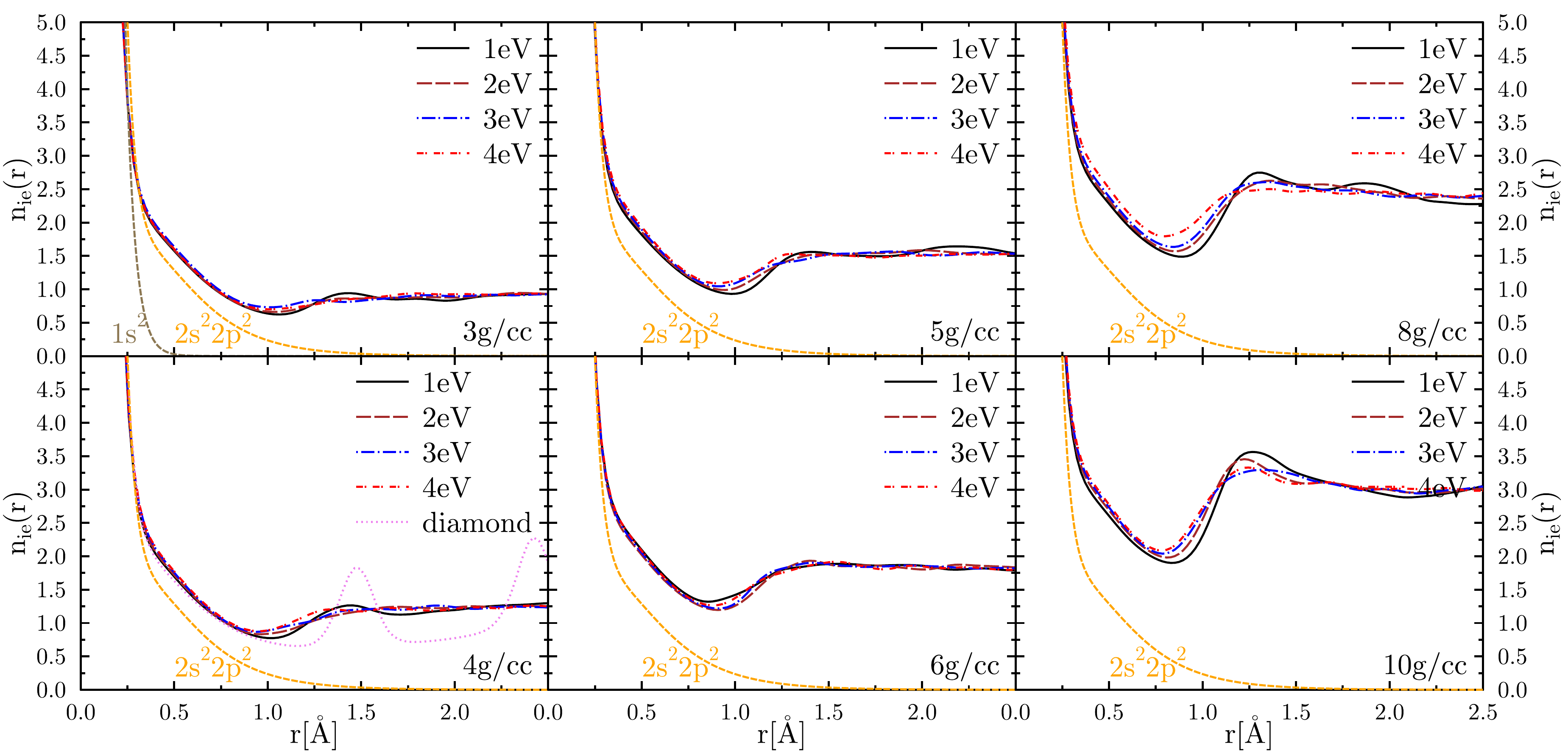}
\caption{Average electron density distribution around an ion for different densities and temperatures of the warm dense carbon fluid. In addition, ideal (atomic) density distributions are shown as well as an example for a diamond case for $4$ g/cc.}
\label{fig:gei}
\end{center}
\end{figure*}
\begin{figure*}[t]
\begin{center}
% reprint
%\includegraphics[width=\textwidth]{elf_rd.pdf}
\includegraphics[width=\textwidth]{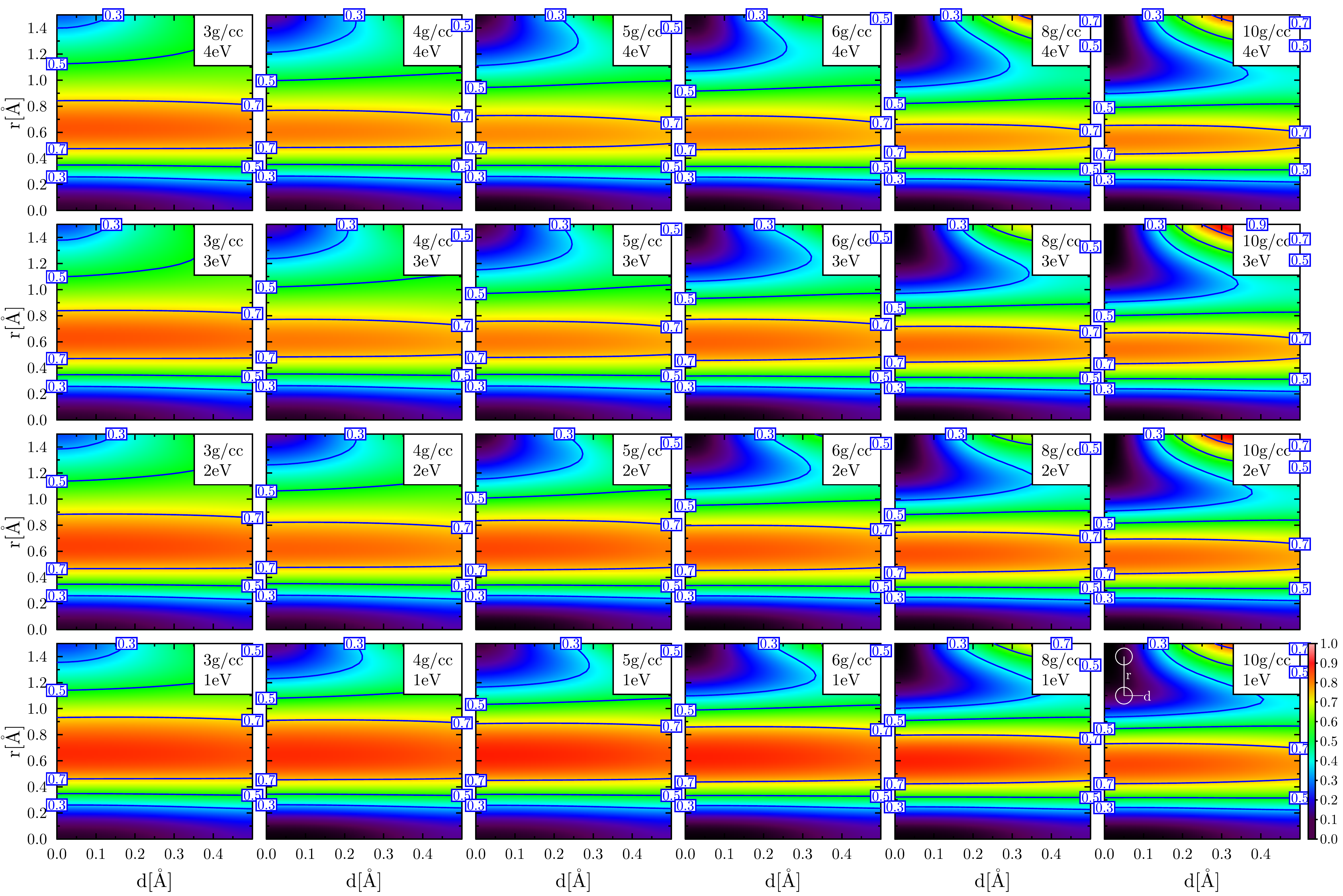}
\caption{Electron localization function for different densities and temperatures of the warm dense carbon fluid. The $r$ axis gives the direction and distance to the nearest neighbor, the $d$ axis is an angular average of the cylinder with radius $d$ perpendicular to $\vec{r}$.}
\label{fig:elf}
\end{center}
\end{figure*}

The clusters naturally have their origin in the covalent bonds between the carbon atoms. Therefore, it is instructive to look at the electron-ion distribution, see Fig. \ref{fig:gei}, and the electron localization function (ELF), see Fig. \ref{fig:elf}. The electron-ion distributions show the increasing overlap of the $n=2$ orbitals but also the continued stable characteristic signature of covalent bonds. The temperature dependence of $g_{ei}$ for densities up to $6$ g/cc is non-existent and very small for higher densities.

%\subsection{Electron localization functions}
A similar systematic picture is provided by the ELF in Fig. \ref{fig:elf}. Here, the maximum of the ELF at half the bond distance is a strong indication for the paired electrons delivering the covalent bond. With increasing temperature and with increasing density, the maximum of the ELF is slightly reduced, showing more detail than $g_{ei}$ or the cluster analysis. Still, stable covalent bonds remain.

\subsection{Electronic density of states \& conductivities}
\begin{figure*}[t]
\begin{center}
% reprint
%\includegraphics[width=\textwidth]{dos_carbon_paper.pdf}
\includegraphics[width=\textwidth]{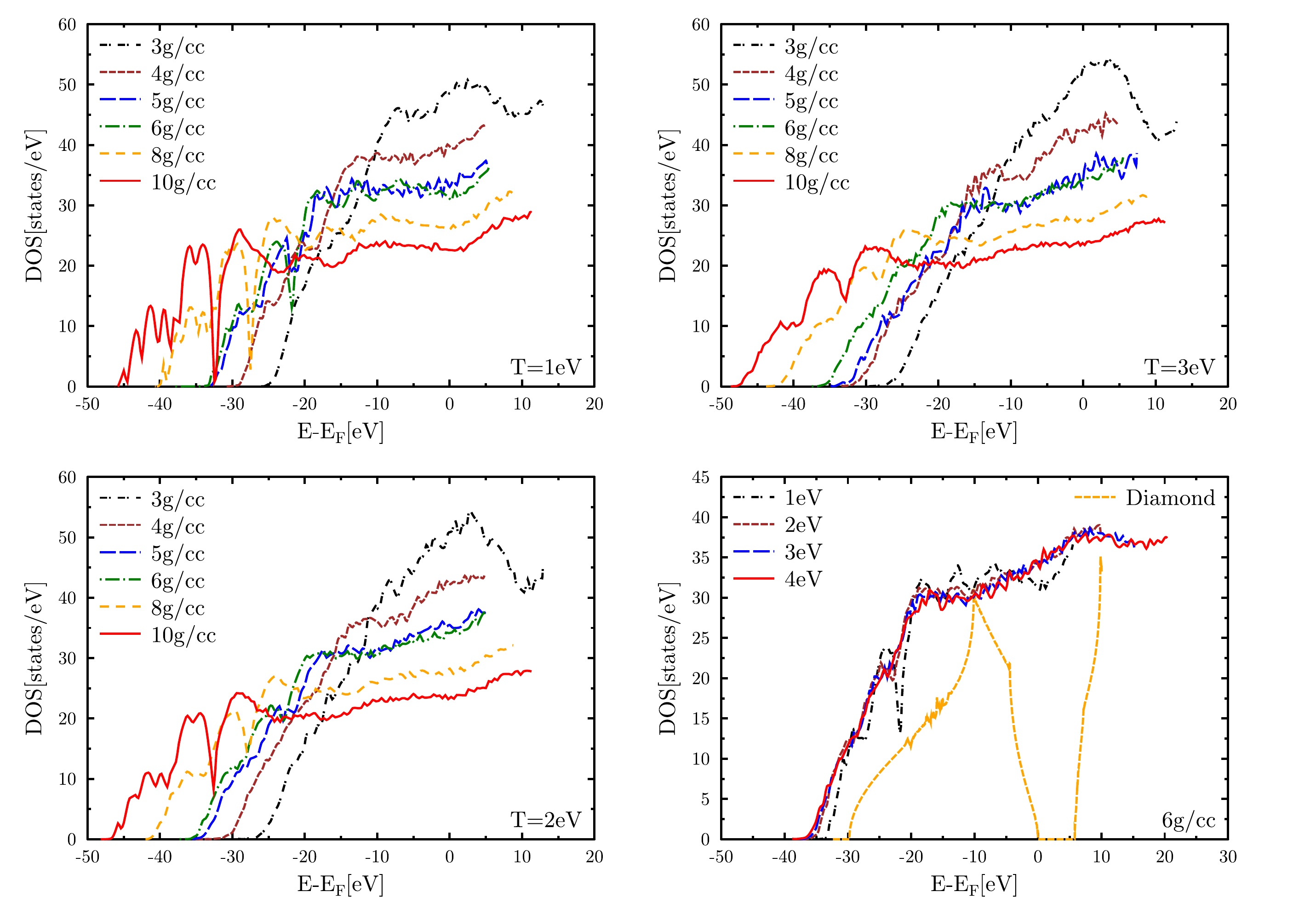}
\caption{The electronic density of states for the warm dense carbon fluid. The bottom right panel shows the change with temperature at a density of $6$ g/cc, all the other panels display a change with density at different constant temperatures.}
\label{fig:dos}
\end{center}
\end{figure*}
In energy space, the eigenstates lie dense, there is no band gap in the parameter range of interest. As a fluid dominated by covalent bonds is usually assumed to possess a band gap, this is in contrast to the picture obtained from the structural analysis. The width of the valence band is increasing with density but only slightly influenced by the temperature. As the absolute value of the DOS is reduced with density, the influence of the density increase is to widen the gaps between the eigenvalues.

%\subsection{Electron conductivities}
\begin{figure*}[t]
\begin{center}
% reprint
%\includegraphics[width=\textwidth]{cond.pdf}
\includegraphics[width=\textwidth]{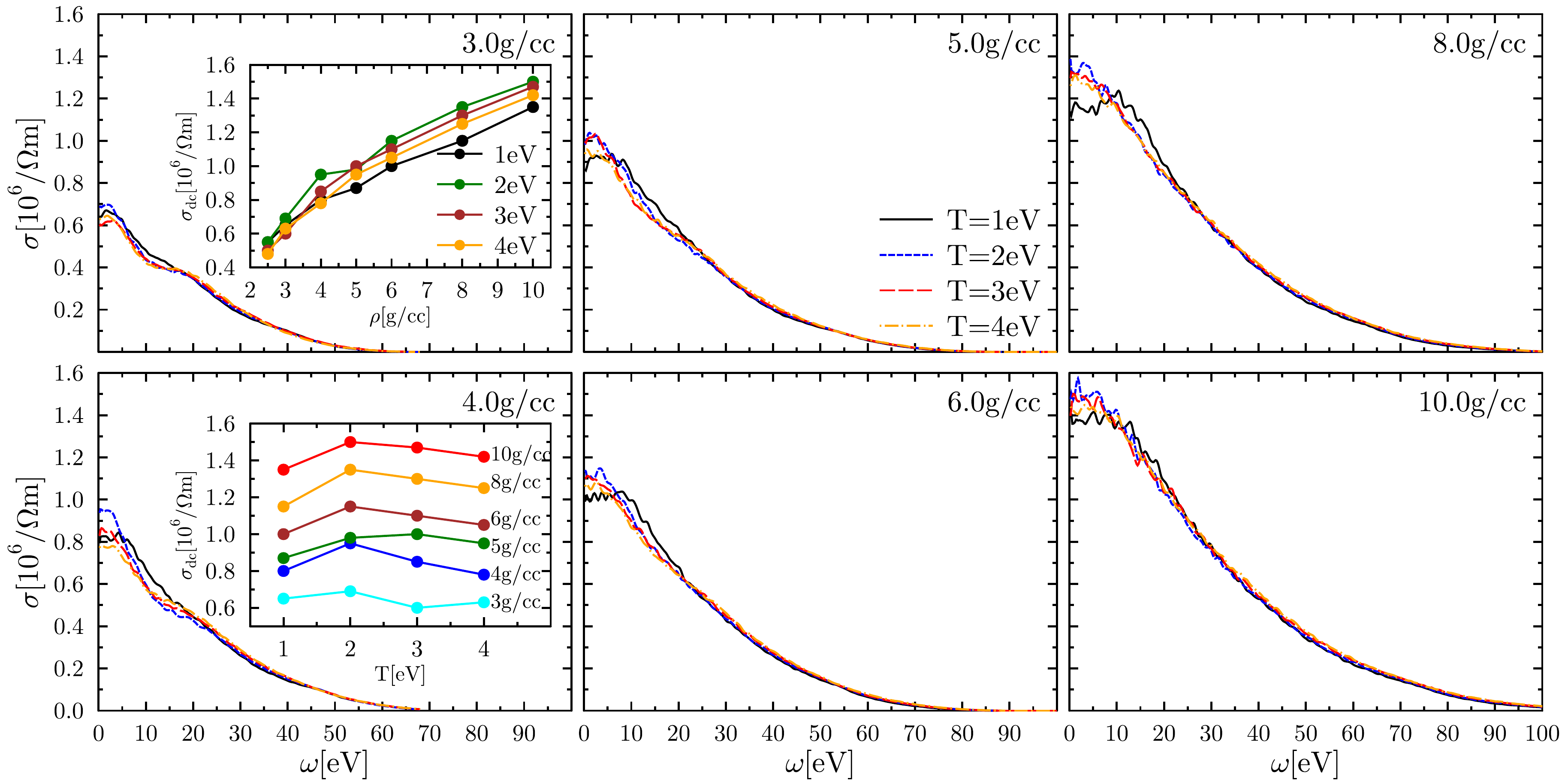}
\caption{The AC and DC electronic conductivity for different densities and temperatures of the warm dense carbon fluid.}
\label{fig:cond}
\end{center}
\end{figure*}
Figure \ref{fig:cond} shows the conductivities in the parameter range of interest. As before, the temperature has little influence for each density. The DC conductivity rises with increasing density and reaches values of a typical metallic fluid. The frequency dependence of the AC conductivity cannot be well approximated by a Drude formula.
%%%%%%%%%%%%%%%%%%%%%%%%%%%%%%%%%%%%%%%%%%%%%%%%%%%%
\subsection{Phase diagram of the high-pressure fluid}
\begin{figure*}[t]
\begin{center}
\includegraphics[width=\columnwidth]{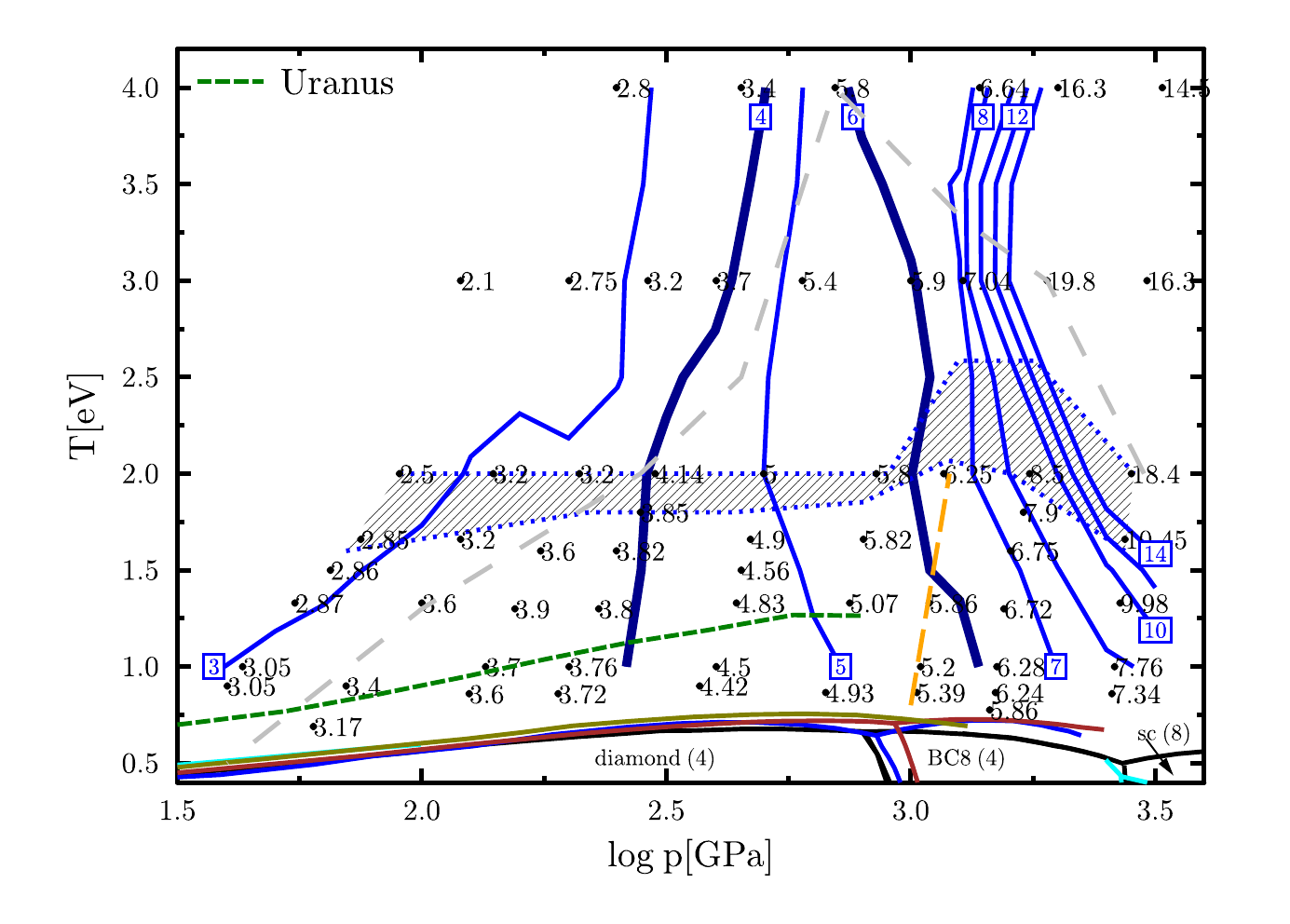}
\caption{Phase space diagram of fluid carbon with coordination numbers given. The solid-fluid  and solid-solid phase boundaries are taken from Ref. \cite{benedict:2014}. The grey shaded area with blue boundaries indicates the parameter space where the double peak in the static structure vanishes. The grey dashed lines indicate the change-over in coordination in the liquid. The orange dashed line indicates where both peaks in the static structure factor have equal height. The blue (thick and thin) lines are contour lines of the coordination numbers. The green dashed line is a possible Uranus isentrope \cite{Nettelmann:2016}.}
\label{fig:pt_coord}
\end{center}
\end{figure*}
Starting from this reasoning, we show the phase space diagram of fluid carbon in Fig. \ref{fig:pt_coord}. The equation of state calculations are in good agreement with the latest equation of state data published by Benedict {\em et al.} \cite{benedict:2014} and do not show any evidence for first order liquid-liquid phase transitions in this pressure-temperature regime. All the indicated structural transitions are of continuous nature. 

The change over of the global maximum in the static structure factor from first peak to second peak is indicated by the dashed orange line in Fig. \ref{fig:pt_coord}. It coincides nicely in pressure with the predicted diamond-BC8 solid-solid phase transition. As stated before, the two main fluid peaks are remnants of the Bragg peaks of the solid phases at similar pressures. The fluid and solid structures are therefore closely related.

The vanishing of the double peak structure in the static structure factor is displayed by a grey shaded area due to a lack of finer resolution in temperature space and due to the difficulty in deciding whether or not a double peak structure exists in slightly noisy data. This transition seems to be independent of the pressure, therefore the phenomena is entirely temperature driven. Beyond a temperature of $T=2\ev$, remaining structural similarities between fluid and solid vanish.

A very interesting behavior of the liquid can be observed when studying the coordination numbers. In the two solid phases of interest for the pressures considered here, the coordination number is fixed at $N_C=4$. In addition, we know from first principle simulations that the melting lines of the diamond and BC8 phases show negative slope, see Fig. \ref{fig:pt_coord}. This means that the fluid needs to have a more efficient packing from $p=5\Mbar$ onward. An indicator for this is the coordination number of the liquid. Following the trend of the fluid coordination number along the $T=1\ev$ isotherm, one observes that  the coordination number has graphite like values of $N_C=3$ for small pressures but finds a steady increase of coordination up to $N_C=8$  for the highest pressures shown. Here, the liquid follows and anticipates the coordination of the diamond, BC8 and sc solid phases. However, for higher temperatures, the liquid is not bound by the restrictions of a solid. For instance, following the $\rho=2.5\gcc$ isochore (leftmost points in p-T space), the coordination drops rapidly as expected for a liquid. However, for most of the pressure-temperature space shown in Fig. \ref{fig:pt_coord}, the coordination number behaves in a non-standard way. 
Firstly, coordination numbers increase with temperature at constant pressure. In a 'normal' liquid, one would expect the numbers to decrease with increasing temperature. The area in which this behavior can be observed is indicated by a wide-spaced dashed grey line. Secondly, very high coordination numbers of beyond $N_c=14$ can be reached. This is due to the peculiar behavior of the pair correlation functions, as shown in Fig. \ref{fig:giii}, which feature the first minimum at particularly large distances for $T\ge 2\ev$ and high densities.
%%%%%%%%%%%%%%%%%%%%%%%%%%%%%%%%%%%%%%%%%%%%%%%%%%%%%%%%%%%%%%%%%%%%%%%%%%%%%%%%
% Summary
%%%%%%%%%%%%%%%%%%%%%%%%%%%%%%%%%%%%%%%%%%%%%%%%%%%%%%%%%%%%%%%%%%%%%%%%%%%%%%%%
\section{Summary}
We have investigated the structure of fluid carbon for densities and pressures corresponding to the graphite, diamond, BC8, and sc solid phases for temperatures up to $4\ev$. The fluid structure shows similarities to the solid structure at corresponding pressure. However, the fluid is more flexible then the solid in adjusting to an increase in pressure and therefore anticipates structural changes (coordination number, packing etc.) at lower pressures than the solid. This leads to two melting line maxima for the diamond and BC8 solid phases as in both cases the fluid shows more efficient packing at high pressures.

The structure of the fluid is dominated by transient carbon-carbon bonds allowing a multitude of clusters of varying sizes to develop and dissociate. This makes the short range order of the carbon fluid extremely rich in features and very distinct from warm dense matter dominated by long range Coulomb forces. In particular, the double peak structure in the static structure factor with changing dominance of first and second peak with pressure is striking. Further, due to relatively high carbon bond dissociation energies, the structure of the fluid remains dominated by such bonds within the investigated temperature regime. Except for the highest temperatures at $2.5\gcc$, ionization stays very low below a temperature of $4\ev$.

These structural properties will be useful in future for laboratory astrophysics experiments investigating the material in the interior of carbon bearing planets \cite{Madhusudhan:2012,Nettelmann:2013,Nettelmann:2016} and for studies on pure carbon using x-ray scattering \cite{kraus:2013,Kraus:2016}.
%%%%%%%%%%%%%%%%%%%%%%%%%%%%%%%%%%%%%%%%%%%%%%%%%%%%%%%%%%%%%%%%%%%%%%%%%%%%%%%%
% Acknowledgments
%%%%%%%%%%%%%%%%%%%%%%%%%%%%%%%%%%%%%%%%%%%%%%%%%%%%%%%%%%%%%%%%%%%%%%%%%%%%%%%%
\section*{Acknowledgments}
KUP and RR acknowledge support of the DFG within the SFB 652. JV acknowledges generous hospitality at the Institut f\"ur Physik of the Universit\"at Rostock.
%\end{acknowledgments}
%%%%%%%%%%%%%%%%%%%%%%%%%%%%%%%%%%%%%%%%%%%%%%%%%%%%%%%%%%%%%%%%%%%%%%%%%%%%%%%%
% Appendix
%%%%%%%%%%%%%%%%%%%%%%%%%%%%%%%%%%%%%%%%%%%%%%%%%%%%%%%%%%%%%%%%%%%%%%%%%%%%%%%%

%%%%%%%%%%%%%%%%%%%%%%%%%%%%%%%%%%%%%%%%%%%%%%%%%%%%%%%%%%%%%%%%%%%%%%%%%%%%%%%%
% literature
%%%%%%%%%%%%%%%%%%%%%%%%%%%%%%%%%%%%%%%%%%%%%%%%%%%%%%%%%%%%%%%%%%%%%%%%%%%%%%%%
\bibliographystyle{elsarticle-num}
%\bibliography{lit_c_structure}{}

%\newpage

%%%%%%%%%%%%%%%%%%%%%%%%%%%%%%%%%%%%%%%%%%%%%%%%%%%%%%%%%%%%%%%%%%%%%%%%%%%%%%%%
% end of main
%%%%%%%%%%%%%%%%%%%%%%%%%%%%%%%%%%%%%%%%%%%%%%%%%%%%%%%%%%%%%%%%%%%%%%%%%%%%%%%%
\end{document}